\documentclass[conference]{IEEEtran}
\IEEEoverridecommandlockouts
\usepackage{cite}
\usepackage{amsmath,amssymb,amsfonts}
\usepackage{algorithmic}
\usepackage[graphicx]{realboxes}
\usepackage{textcomp}
\usepackage[table,xcdraw]{xcolor}
\usepackage{multirow}
\usepackage{array}
\usepackage{tikz}
\usepackage{siunitx}
\usepackage{lscape}
\usepackage{adjustbox}
\usepackage{tabularx}
\usepackage{mleftright}
\usepackage{etoolbox}

\makeatletter
\newcommand*{\rom}[1]{\expandafter\@slowromancap\romannumeral #1@}
\makeatother

\def\BibTeX{{\rm B\kern-.05em{\sc i\kern-.025em b}\kern-.08em
    T\kern-.1667em\lower.7ex\hbox{E}\kern-.125emX}}
\begin{document}

\title{Assessment of Unconsciousness \\ for Memory Consolidation Using EEG Signals
}

\author{\IEEEauthorblockN{Gi-Hwan Shin}
\IEEEauthorblockA{\textit{Dept. Brain and Cognitive Engineering} \\
\textit{Korea University}\\
Seoul, Republic of Korea \\
gh\_shin@korea.ac.kr}
\and
\IEEEauthorblockN{Minji Lee}
\IEEEauthorblockA{\textit{Dept. Brain and Cognitive Engineering} \\
\textit{Korea University}\\
Seoul, Republic of Korea \\
minjilee@korea.ac.kr}
\and
\IEEEauthorblockN{Seong-Whan Lee}
\IEEEauthorblockA{\textit{Dept. Artificial Intelligence} \\
\textit{Korea University}\\
Seoul, Republic of Korea \\
sw.lee@korea.ac.kr}

\thanks{20xx IEEE. Personal use of this material is permitted. Permission
from IEEE must be obtained for all other uses, in any current or future media, including reprinting/republishing this material for advertising or promotional purposes, creating new collective works, for resale or redistribution to servers or lists, or reuse of any copyrighted component of this work in other works.}

\thanks{This work was supported by the Institute for Information \& Communications Technology Planning \& Evaluation (IITP) grant funded by the Korea government (No. 2017-0-00451, Development of BCI based Brain and Cognitive Computing Technology for Recognizing User’s Intentions using Deep Learning).}
}

\maketitle

\begin{abstract} % 
The assessment of consciousness and unconsciousness is a challenging issue in modern neuroscience. Consciousness is closely related to memory consolidation in that memory is a critical component of conscious experience. So far, many studies have been reported on memory consolidation during consciousness, but there is little research on memory consolidation during unconsciousness. Therefore, we aim to assess the unconsciousness in terms of memory consolidation using electroencephalogram signals. In particular, we used unconscious state during a nap; because sleep is the only state in which consciousness disappears under normal physiological conditions. Seven participants performed two memory tasks (word-pairs and visuo-spatial) before and after the nap to assess the memory consolidation during unconsciousness. As a result, spindle power in central, parietal, occipital regions during unconsciousness was positively correlated with the performance of location memory. With the memory performance, there was also a negative correlation between delta connectivity and word-pairs memory, alpha connectivity and location memory, and spindle connectivity and word-pairs memory. We additionally observed the significant relationship between unconsciousness and brain changes during memory recall before and after the nap. These findings could help present new insights into the assessment of unconsciousness by exploring the relationship with memory consolidation. \\
\end{abstract}


\section{Introduction}
% BMI 설명
Brain-machine interface (BMI) has been widely used to assess the levels of consciousness \cite{pan2014detecting}. In particular, these techniques are helpful to diagnose patients with disorders of consciousness \cite{lule2013probing, zhu2016canonical}. The fact that consciousness consists of wakefulness and awareness is very important in that the standards of consciousness can be different \cite{laureys2005neural}. The wakefulness includes arousal, alertness, and vigilance, while awareness is the sum of cognitive and emotional functions \cite{sander2015assessment}. Simply put, wakefulness indicates whether the user could react to external stimuli and awareness represents whether the user has conscious experience. For example, the ketamine-induced unconsciousness has a dreamlike conscious experience but has no response to external stimuli, so it can be said to be low wakefulness but high awareness \cite{sarasso2015consciousness}.

Many studies have reported that assessment of consciousness using electroencephalogram (EEG), which is low cost and a high temporal resolution \cite{song2019possible, chen2016high, lee2015subject}. From an awareness point of view, delta connectivity and spectral exponent of the resting EEG could evaluate consciousness \cite{colombo2019spectral, lee2019connectivity}. In addition, the difference in brain connectivity between propofol-induced unconsciousness and wakefulness was especially found in the parietal region \cite{lee2017network}. Recently, studies have been reported to assess consciousness from a wakefulness perspective \cite{lendner2019electrophysiological}. However, it is still unclear about the assessment of consciousness in the wakefulness.

%의식과 메모리
Memory is deeply related to consciousness \cite{velichkovsky2017consciousness}. In specific, memory consolidation makes it possible for memories of our daily experiences to be stored in an enduring way during sleep and wakefulness \cite{oudiette2013role}. Therefore, memory performance itself may be used as a measure of consciousness assessment in that memory recall is directly related to the levels of consciousness. Many studies show that changes in spindle band and parietal regions are linked to memory during consciousness \cite{kalafatovich2020neural}. However, research about changes associated with memory consolidation during unconsciousness is to be needed. 

\begin{figure*}[t!]
\centering
\scriptsize
\includegraphics[width=\textwidth]{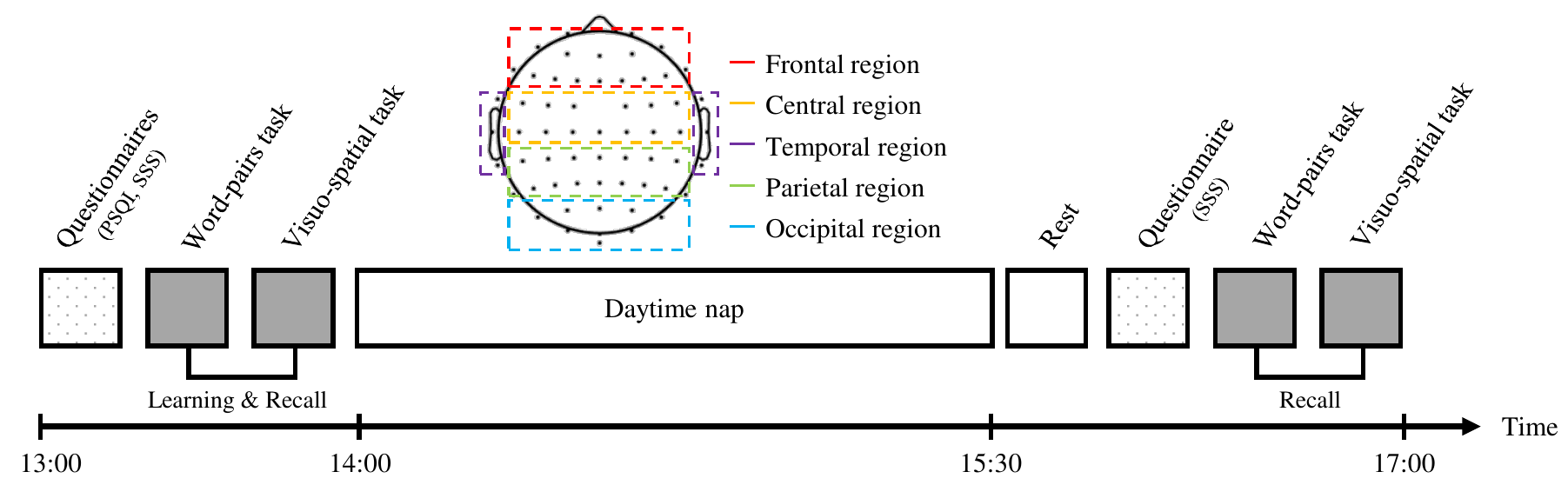}
\caption{Experimental procedures. The experiment consisted of questionnaires, memory tasks, and the nap (unconscious state). 
PSQI and SSS questionnaires are performed before the nap, and only SSS questionnaire is performed after the nap. The memory tasks are performed before and after the nap. PSQI = Pittsburg sleep quality index, SSS = Stanford sleepiness scale.}
\end{figure*}
%%%%%%%%%%%%%%%%%%%%%%%%%%%%%%%%%%%%%%%%%%%%%%%%%%%%%%%%%%%%%%%%%%%%%%%%%%%%%%%%

% In this study~
In this study, we aimed to assess the unconsciousness in terms of memory consolidation using EEG signals. In particular, we focused on wakefulness that could not react to external stimuli, not awareness. Therefore, we used the nap to maintain the unconscious state in normal physiological conditions \cite{siclari2017neural}. We hypothesized that there is the strong relationship between memory consolidation and EEG features during unconsciousness. Our results could present as a new biomarker in assessing unconsciousness in terms of memory consolidation.

\section{Methods}
\subsection{Participants} 
Seven healthy adults (all males, age 26.0 $\pm$ 2.4 years) participated in this study. No participants had any history of neurologic, psychiatric, sleep, or internal disorders. This study was approved by the Institutional Review Board at Korea University (KUIRB-2020-0112-01), and each participant gave written informed consent before the experiments.

\subsection{Experimental Procedure}
All participants visited the laboratory at noon and prepared EEG recordings. They answered the questionnaires about the subjective quality of sleep before the experiment (Pittsburg sleep quality index, PSQI) \cite{buysse1989pittsburgh} and current sleepiness (Stanford sleepiness scale, SSS) \cite{hoddes1972stanford}. The memory tasks consisting of the learning, immediate recall, and delayed recall sessions were performed before the nap. At 2:00 pm, the participants were asked to take the nap for 90 min. After waking up, they took a rest for 30 min and reported the SSS, again. The delayed recall session was performed to investigate the effect on memory consolidation (Fig. 1). 

%Declarative memory refers to the ability to recall past events, facts, and general knowledge and usually includes a delayed retrieval. Procedural memory is a part of the long-term memory that is responsible for knowing how to do things, also known as motor skills.
\subsection{Memory Tasks}
The memory tasks consisted of two declarative memory tests (a word-pairs task \cite{leminen2017enhanced} and a visuo-spatial task including picture and location memory \cite{ladenbauer2016brain}) (Fig. 2). All memory tasks were implemented with Psychtoolbox (http://psychtoolbox.org). Participants were instructed to memorize items for recall of memory, but no specific strategy was recommended \cite{ladenbauer2017promoting}.

\subsubsection{Word-pairs memory} 
The word-pairs task included 108 semantically related word-pairs (e.g.,``event-festival") \cite{marshall2006boosting}. The order of word-pairs was presented randomly for each participant. In the learning session, each pair of words was presented during 4 sec, preceded by an inter-stimulus interval (ISI) that lasted for 1 sec. The immediate recall session was performed after the learning session. The participants were asked to enter a word that corresponds to the word displayed on the screen. After the response, the correct answer was visible for 2 sec independent of the correctness of the participant's answer. After the unconscious state, the delayed recall session was the same except that no correct answer was displayed.
%%%%%%%%%%%%%%%%%%%%%%%%%%%%%%%%%%%%%%%%%%%%%%%%%%%%%%%%%%%%%%%%%%%%%%%%%%%%%%%%
\begin{figure}[t!]
\centering
\scriptsize
\includegraphics[width=\columnwidth]{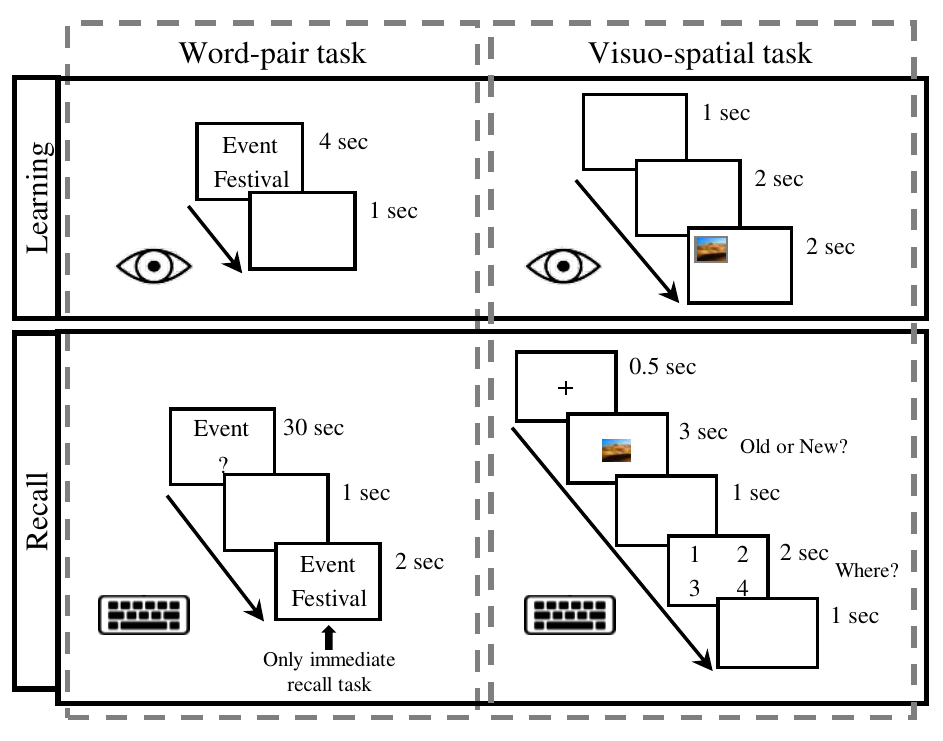}
\caption{A detailed description of the word-pairs and visuo-spatial memory tasks. The word-pairs memory task consisted of 54 trials in both the learning and recall sessions, and the visuo-spaital memory task consisted of 38 trials and 76 trials in the learning and recall sessions, respectively.}
\end{figure}
%%%%%%%%%%%%%%%%%%%%%%%%%%%%%%%%%%%%%%%%%%%%%%%%%%%%%%%%%%%%%%%%%%%%%%%%%%%%%%%%

The memory performance was evaluated by summing all correctly recalled words. The correct answers included responses with typos or inflectional error. In addition, derivations mistakes were counted as errors.

\subsubsection{Visuo-spaital memory}
The visuo-spatial memory task required participants to learn 38 neutral pictures (objects, scenes taken from the SUN database\cite{xiao2010sun}) (picture memory) and additionally to memorize the location at which they were presented (location memory). During each trial, a fixation cross was first presented during 1 sec, preceded by a gray square randomly at one of the 4 quadrants on the screen for 2 sec. This was followed by a picture within the square for 2 sec. The immediate and delayed recall sessions showed a fixation cross for 1 sec and subsequent stimuli (38 learned and 38 new pictures) were randomly displayed in the center of the screen for 3 sec. Within this period, participants pressed old (``o") or new (``n") buttons (picture memory). If they recognized a picture, they also selected in which quadrant they believed the picture had been presented (location memory).

As a measure of picture recognition memory performance, percent correct responses were determined for each participant as follows: the proportion of correct old responses + proportion of correct new responses. To determine the performance of location memory, both correctly and incorrectly recall picture locations were taken into account as follows: number of correctly recall/number of correct old responses - number of falsely recall location/number of correct old responses.

\subsection{Data Acquisition and Preprocessing}
The EEG signals were recorded at 1,000 Hz using an amplifier (BrainAmp; Brain Project GmBH, Germany). The 60 Ag/AgCl electrodes were used according to the 10-20 international system. Additionally, the reference electrode was located at FCz and the reference electrode was placed in AFz. For all electrodes impedance was kept below 10 k$\Omega$. 

The EEG signals were processed with MATLAB R2018b using the EEGLAB toolbox \cite{delorme2004eeglab}. Data were down-sampled to 250 Hz and band-pass filtered between 0.5 to 50 Hz. We divided into nap and memory recall. EEG data of the 15 min before and after were excluded during the 90 min nap to clearly include the non-responsive state in external stimuli. Then the 60 min data was segmented into 3 sec intervals. In the memory recall, we segmented the EEG signals at the specific time that is most prominent when perceived by the brain for each task (word-pairs memory: 400-800 msec \cite{bader2010recognition}, picture and location memory: 200-400 msec \cite{schneider2016time}). To remove artifacts such as muscle movements and eye blinks, the epochs were excluded when the amplitude value exceeded $\pm$ 100 \si{\micro}V. The independent component analysis was also performed to remove components with dominant artifacts in the memory task.

\subsection{Data Analysis} 

To compare the EEG feature in the spectral and spatial domain, we grouped into five brain regions as follows: frontal, central, temporal, parietal, and occipital regions. In addition, we divided into six frequency bands as follows: delta (0.5-4 Hz), theta (4-7 Hz), alpha (7-12 Hz), spindle (12-16 Hz), beta (16-30 Hz), and gamma (30-50 Hz) bands. 

\subsubsection{Spectral power} 
The signals were processed to analyze EEG characteristics in the frequency domain using the fast Fourier transform (FFT). The power spectral density (PSD) was calculated for each frequency component composing those EEG signals \cite{suk2012novel}:

\begin{equation} PSD_{f_1-f_2} = 10*log_{10}(2\int_{f_1}^{f_2} |\hat{x}(2\pi f)|^2 df) \end{equation}

\noindent
where $f_1$, $f_2$ represent the lower and upper frequencies respectively, and $\hat{x}(2\pi f)$ was obtained by FFT. $10*log_{10}(\bullet)$ denotes unit conversion from microvolts to decibels.

\subsubsection{Functional connectivity}
To investigate the functional connectivity among brain regions \cite{ding2013changes}, we used the weighted phase lag index (wPLI). This measure is used to identify non-zero phase lag statistical inter-dependencies between EEG time series from pairs of channels. Specifically, wPLI was calculated to minimize the impact of volume conduction and the number of artifacts \cite{lee2017change}:

\begin{equation} wPLI = \frac{|E\{\mathcal{J}\{X\}\}|}{E\{|\mathcal{J}\{X\}|\}} =  \frac{|E\{|\mathcal{J}\{X\}|sgn(\mathcal{J}\{X\})\}|}{E\{|\mathcal{J}\{X\}|\}}  \end{equation}

\noindent
where $\mathcal{J}\{X\}$ is based only on the imaginary component of the cross-spectrum $X=Z_i Z_j^*$ between two channels, $Z_i$ is the complex-valued Fourier transform of the signal of channel $i$, $Z_j^*$ is the complex conjugate of $Z_j$, and $E\{\bullet\}$ means the expected-value operator.

\subsection{Statistical Analysis} 
We performed a non-parametric paired permutation test (\textit{r} = 1,000) to compare EEG features during unconsciousness and memory-related characteristics. We also used Pearson's correlation to examine the relationship between differences in memory task performance and the EEG features during unconsciousness. In addition, the Kruskal-Wallis test (non-parametric one way analysis of variance) was performed to investigate the neurophysiological changes associated with memory before and after the nap. For post-hoc analysis, a paired permutation was used with Bonferroni's correction (\textit{r} = 1,000). The alpha level was set at 0.05 for all statistical significance.

%Performance of each memory task before and after the nap.
%%%%%%%%%%%%%%%%%%%%%%%%%%%%%%%%%%%%%%%%%%%%%%%%%%%%%%%%%%%%%%%%%%%%%%%%%%%%%%%%
\begin{figure}[t!]
\centering
\scriptsize
\includegraphics[width=\columnwidth]{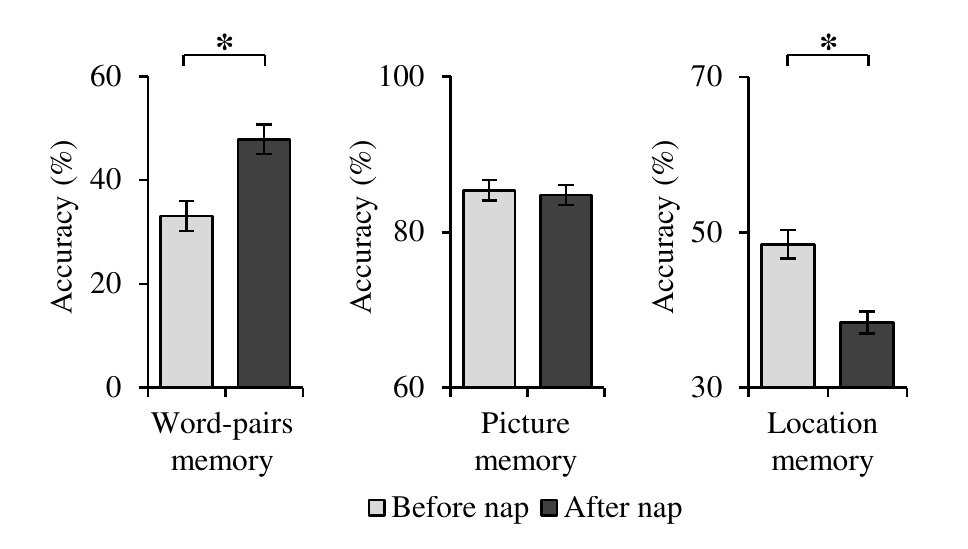}
\caption{Memory task performance. Error bars show standard errors. * indicates statistical significance at \textit{p} $<$ 0.05 determined by paired permutation test.}
\end{figure}
%%%%%%%%%%%%%%%%%%%%%%%%%%%%%%%%%%%%%%%%%%%%%%%%%%%%%%%%%%%%%%%%%%%%%%%%%%%%%%%%

\section{Results}
\subsection{Nap Quality and Memory Task Performance} %fig 3 
All participants had no sleep problems (PSQI $>$ 8). In addition, for nap quality, there were no statistical differences in SSS before and after the nap (\textit{t} = -2.121, \textit{p} = 0.153). Therefore, memory task itself before the nap did not affect nap quality.
%%%%%%%%%%%%%%%%%%%%%%%%%%%%%%%%%%%%%%%%%%%%%%%%%%%%%%%%%%%%%%%%%%%%%%%%%%%%%%%%
\begin{figure*}[t!]
\centering
\scriptsize
\includegraphics[width=\textwidth]{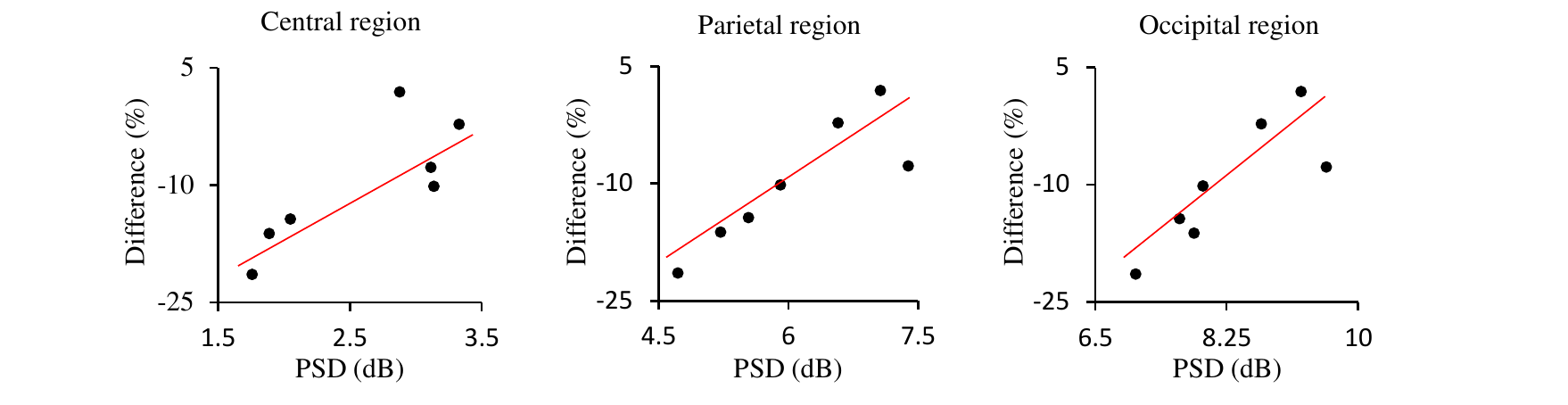}
\caption{Correlation between spindle PSD during unconsciousness and difference in the location memory performance. The differences in memory performance indicate the performance between immediate and delayed memory recall sessions. PSD = power spectral density.}
\end{figure*}
%%%%%%%%%%%%%%%%%%%%%%%%%%%%%%%%%%%%%%%%%%%%%%%%%%%%%%%%%%%%%%%%%%%%%%%%%%%%%%%%
%%%%%%%%%%%%%%%%%%%%%%%%%%%%%%%%%%%%%%%%%%%%%%%%%%%%%%%%%%%%%%%%%%%%%%%%%%%%%%%%
\begin{figure*}[t!]
\centering
\scriptsize
\includegraphics[width=\textwidth]{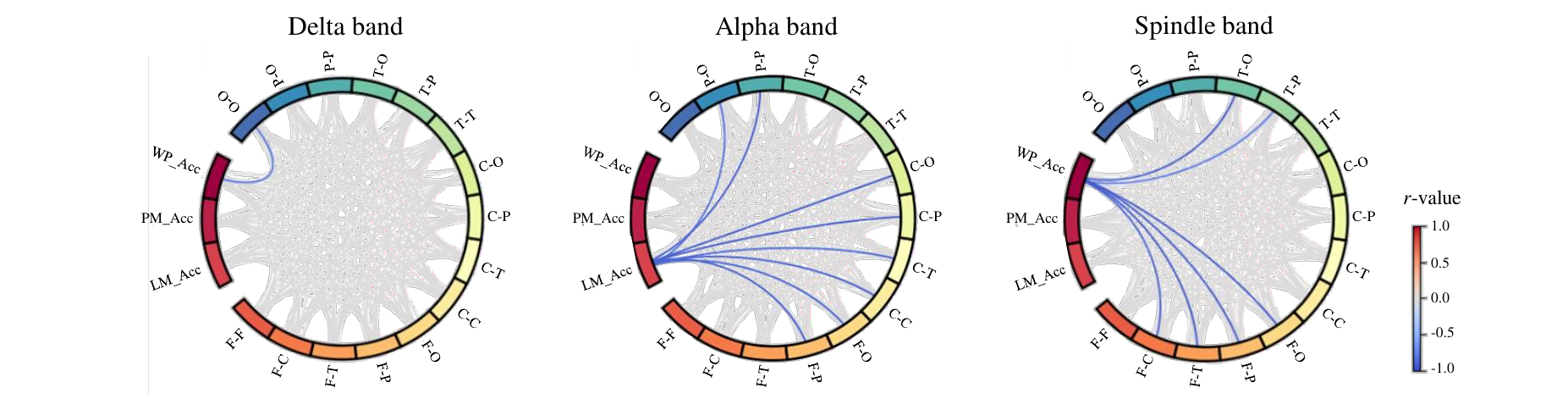}
\caption{Correlation between wPLI during unconsciousness and difference in the memory performance. Blue lines indicate the significantly negative correlation, whereas gray lines indicate no significant connections. wPLI = weighted phase lag index, WP\_Acc = difference in word-pairs memory performance between immediate and delayed memory recall sessions, PM\_Acc = difference in picture memory performance between immediate and delayed memory recall sessions, LM\_Acc = difference in location memory performance between immediate and delayed memory recall sessions, F = frontal region, C = central region, T = temporal region, P = parietal region, O = occipital region.}
\end{figure*}
%%%%%%%%%%%%%%%%%%%%%%%%%%%%%%%%%%%%%%%%%%%%%%%%%%%%%%%%%%%%%%%%%%%%%%%%%%%%%%%%
Fig. 3 showed the average performance of all participants in each memory task. As a result, word-pairs memory was significantly increased (\textit{t} = 6.940, \textit{p} = 0.021), and the location memory was significantly decreased (\textit{t} = -0.647, \textit{p} = 0.023) after the nap. On the other hand, there were no significant differences in picture memory (\textit{t} = -0.276, \textit{p} = 0.642). 

\subsection{Assessment of Unconsciousness with Memory Performance} %fig 4
We examined the relationship between PSD during unconsciousness and differences in the memory performance. Fig. 4 showed a strong positive correlation between the performance of location memory and PSD in spindle band (central region (\textit{r} = 0.813, \textit{p} = 0.026); parietal region (\textit{r} = 0.869, \textit{p} = 0.011); occipital region (\textit{r} = 0.839, \textit{p} = 0.018) (Fig. 4). There was no correlation between PSD in other frequency bands and location memory. In addition, we observed that the word-pairs and picture memory have no correlation with PSD during unconsciousness, respectively.

%fig5
Fig. 5 showed the relationship between wPLI during unconsciousness and difference in memory performance. Unlike PSD, negative correlation was observed in wPLI. Specifically, negative relationship with word-pairs performance was found in the occipital wPLI (\textit{r} = -0.773, \textit{p} = 0.042) in the delta band. We also explored a significant negative correlation with the performance of location memory in alpha band (frontal-parietal wPLI: \textit{r}-value = -0.872, \textit{p} = 0.011;  frontal-occipital wPLI: \textit{r} = -0.845, \textit{p} = 0.017; central wPLI: \textit{r} = -0.863, \textit{p} = 0.012; central-temporal wPLI: \textit{r} = -0.815, \textit{p} = 0.026; central-parietal wPLI: \textit{r} = -0.886, \textit{p} = 0.008; central-occipital wPLI: \textit{r} = -0.876, \textit{p} = 0.010; parietal-parietal wPLI: \textit{r} = -0.870, \textit{p} = 0.011; parietal-occipital wPLI: \textit{r} = -0.778, \textit{p} = 0.039). Similarly, in the spindle band, there was a negative correlation between the performance of the word-pairs memory (frontal-central wPLI: \textit{r} = -0.859, \textit{p} = 0.013; frontal-temporal wPLI: \textit{r} = -0.921, \textit{p} = 0.003; frontal-parietal wPLI: \textit{r} = -0.903, \textit{p} = 0.005; frontal-occipital wPLI: \textit{r} = -0.916, \textit{p} = 0.004; temporal-parietal wPLI: \textit{r} = -0.756, \textit{p} = 0.049) temporal-occipital wPLI: \textit{r} = -0.897, \textit{p} = 0.006). On the other hand, wPLI in the theta, beta, gamma bands showed no significant correlation with memory task performance.

%%%%%%%%%%%%%%%%%%%%%%%%%%%%%%%%%%%%%%%%%%%%%%%%%%%%%%%%%%%%%%%%%%%%%%%%%%%%%%%%
\begin{figure*}[t!]
\centering
\scriptsize
\includegraphics[width=\textwidth]{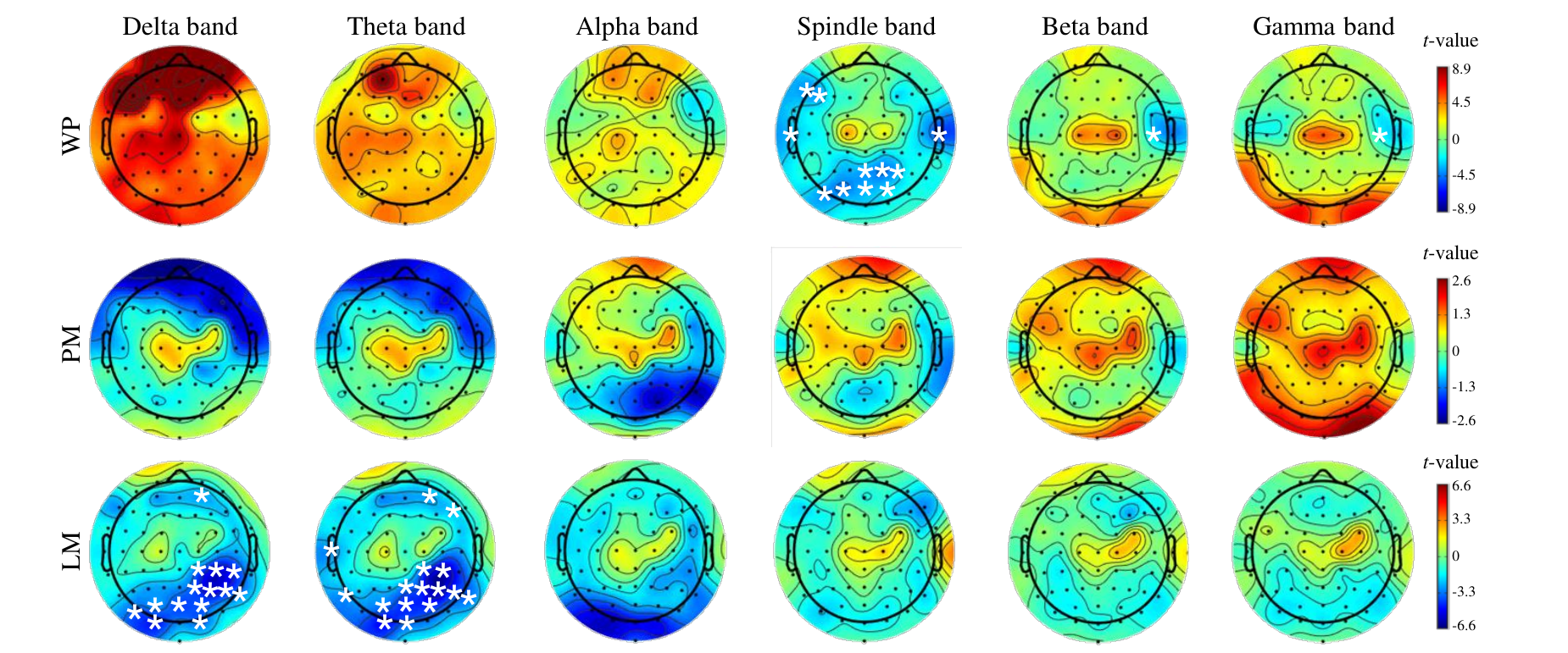}
\caption{Statistical differences in PSD during successful memory recall before and after the nap. The white asterisk indicates a significant electrode in spectral power (\textit{p} $<$ 0.05 with Bonferroni's correction). PSD = power spectral density, WP = word-pairs memory, PM = picture memory, LM = location memory.}
\end{figure*}
%%%%%%%%%%%%%%%%%%%%%%%%%%%%%%%%%%%%%%%%%%%%%%%%%%%%%%%%%%%%%%%%%%%%%%%%%%%%%%%%

\subsection{Difference in EEG Features According to Nap during Memory Recall } %(Fig. 6) 
%% PSD 낮잠 전후 통계적 비교 (기억)
We observed the spatial differences in spectral power during memory recall before and after the nap (Fig. 6). In the word-pairs memory decreased spindle PSD in the temporal and occipital regions during recall task after the unconscious. In the location memory, delta and theta PSD in the parietal and occipital regions decreased during recall tasks after the unconscious. On the other hand, there was no spatial difference in picture memory during memory recall.

%% wPLI 낮잠 전후 통계적 비교 (기억)
Additionally, differences in wPLI during memory recall were explored. In the word pair memory, all the significant functional connectivity of the delayed recall after unconsciousness increased in the delta band (frontal wPLI: \textit{t} = 3.461, \textit{p} = 0.016; frontal-central wPLI: \textit{t} = 2.431, \textit{p} = 0.047), theta band (frontal wPLI: \textit{t} = 2.498, \textit{p} = 0.046) and gamma band (central-temporal wPLI: \textit{t} = 3.395, \textit{p} = 0.015; temporal wPLI: \textit{t} = 2.598, \textit{p} $<$ 0.001; temporal-parietal wPLI: \textit{t} = 2.161, \textit{p} = 0.046; parietal wPLI: \textit{t} = -2.710, \textit{p} = 0.047). In addition, picture memory showed a significant difference only in the alpha band. In specific, the frontal wPLI increased during recall task after unconsciousness (\textit{t} = 2.359, \textit{p} = 0.046) and central-parietal wPLI decreased during recall task after unconsciousness (\textit{t} = -2.984, \textit{p} = 0.015). The location memory showed a significant difference only in theta band when comparing immediate and delayed recall sessions, and wPLI decreased after unconsciousness (central wPLI: \textit{t} = -2.352, \textit{p} = 0.030; parietal wPLI: \textit{t} = -3.088, \textit{p} = 0.032).
%%%%%%%%%%%%%%%%%%%%%%%%%%%%%%%%%%%%%%%%%%%%%%%%%%%%%%%%%%%%%%%%%%%%%%%%%
\begin{table*}[t!]
\caption{Correlation between Gamma PSD during Unconsciousness and Differences in Gamma PSD during Memory Recall Before and After the Nap}
\centering
\small
\renewcommand{\arraystretch}{1.2}
\begin{tabular*}{\textwidth}{@{\extracolsep{\fill}\quad}lcccccc}
\hline
{\color[HTML]{000000} }                 & \multicolumn{2}{c}{{\color[HTML]{000000} \textbf{Word-pairs memory}}}                                & \multicolumn{2}{c}{{\color[HTML]{000000} \textbf{Picture memory}}}                                  & \multicolumn{2}{c}{{\color[HTML]{000000} \textbf{Location   memory}}}                               \\ \cline{2-7}
{\color[HTML]{000000} }                 & {\color[HTML]{000000} \textit{\textbf{r-value}}} & {\color[HTML]{000000} \textit{\textbf{p-value}}} & {\color[HTML]{000000} \textit{\textbf{r-value}}} & {\color[HTML]{000000} \textit{\textbf{p-value}}} & {\color[HTML]{000000} \textit{\textbf{r-value}}} & {\color[HTML]{000000} \textit{\textbf{p-value}}} \\ \hline
{\color[HTML]{000000} Frontal region}   & {\color[HTML]{000000} 0.869}                     & {\color[HTML]{000000} \textit{0.011}}            & {\color[HTML]{000000} 0.962}                     & {\color[HTML]{000000} \textit{\textless{}0.001}} & {\color[HTML]{000000} 0.817}                     & {\color[HTML]{000000} \textit{0.025}}            \\
{\color[HTML]{000000} Central region}   & {\color[HTML]{000000} 0.441}                     & {\color[HTML]{000000} 0.322}                     & {\color[HTML]{000000} 0.803}                     & {\color[HTML]{000000} \textit{0.029}}            & {\color[HTML]{000000} 0.727}                     & {\color[HTML]{000000} 0.064}                     \\
{\color[HTML]{000000} Temporal region}  & {\color[HTML]{000000} 0.038}                     & {\color[HTML]{000000} 0.936}                     & {\color[HTML]{000000} 0.649}                     & {\color[HTML]{000000} 0.115}                     & {\color[HTML]{000000} 0.504}                     & {\color[HTML]{000000} 0.249}                     \\
{\color[HTML]{000000} Parietal region}  & {\color[HTML]{000000} 0.776}                     & {\color[HTML]{000000} \textit{0.040}}            & {\color[HTML]{000000} 0.849}                     & {\color[HTML]{000000} \textit{0.016}}            & {\color[HTML]{000000} 0.816}                     & {\color[HTML]{000000} \textit{0.025}}            \\
{\color[HTML]{000000} Occipital region} & {\color[HTML]{000000} 0.796}                     & {\color[HTML]{000000} \textit{0.032}}            & {\color[HTML]{000000} 0.929}                     & {\color[HTML]{000000} \textit{0.002}}            & {\color[HTML]{000000} 0.956}                     & {\color[HTML]{000000} \textit{\textless{}0.001}} \\ \hline
\end{tabular*}
\end{table*}
%%%%%%%%%%%%%%%%%%%%%%%%%%%%%%%%%%%%%%%%%%%%%%%%%%%%%%%%%%%%%%%%%%%%%%%%%%%%
%%%%%%%%%%%%%%%%%%%%%%%%%%%%%%%%%%%%%%%%%%%%%%%%%%%%%%%%%%%%%%%%%%%%%%%%%%%%%%%%
\begin{figure*}[t!]
\centering
\scriptsize
\includegraphics[width=\textwidth]{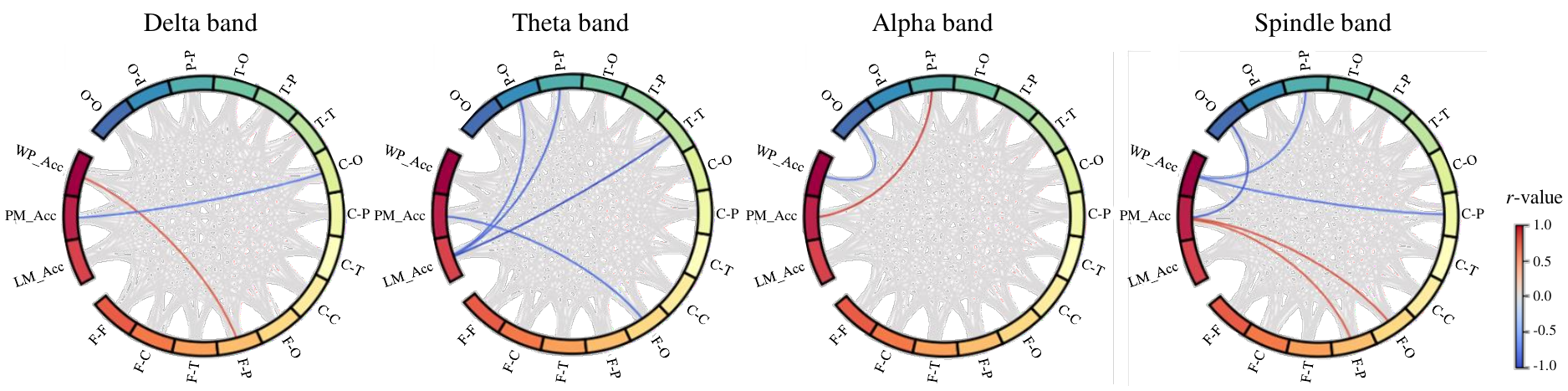}
\caption{Correlation between wPLI and difference in memory performance. Red, blue, and gray lines indicate the significantly positive correlation, the significantly negative correlation, and no significant connections, respectively. wPLI = weighted phase lag index, WP\_Acc = difference in word-pairs memory performance between immediate and delayed memory recall sessions, PM\_Acc = difference in picture memory performance between immediate and delayed memory recall sessions, LM\_Acc = difference in location memory performance between immediate and delayed memory recall sessions, F = frontal region, C = central region, T = temporal region, P = parietal region, O = occipital region.}
\end{figure*}
%%%%%%%%%%%%%%%%%%%%%%%%%%%%%%%%%%%%%%%%%%%%%%%%%%%%%%%%%%%%%%%%%%%%%%%%%%%%%%%%
%
\subsection{Assessment of Unconsciousness with EEG Features during Memory Recall} %Table 1, Fig. 7
To investigate the assessment of unconsciousness with differences in PSD during memory recall before and after nap, Table \rom{1} showed a strong positive relationship in PSD during unconsciousness and memory recall in the gamma band. In addition, word-pairs memory indicated positive relationship PSD in alpha band (central region: \textit{r} = 0.777, \textit{p} = 0.039; temporal region: \textit{r} = 0.775, \textit{p} = 0.041), spindle band (\textit{r} = 0.839, \textit{p} = 0.018), and beta band (\textit{r} = 0.832, \textit{p} = 0.002). The picture memory showed a significant positive correlation PSD in only theta band (temporal region: \textit{r} = 0.802, \textit{p} = 0.029), and the location memory showed a significant positive correlation PSD in only alpha band (central region: \textit{r} = 0.807, \textit{p} = 0.028; occipital region: \textit{r} = 0.756, \textit{p} = 0.049). In summary, gamma PSD has mainly a positive correlation between unconsciousness and differences in memory recall.

Fig. 7 indicated the important relationship with wPLI during memory recall. In the delta band, a positive correlation was found between the word-pairs memory and the frontal-parietal wPLI (\textit{r} = 0.757, \textit{p} = 0.049). We additionally observed a negative correlation between picture memory and central-occipital wPLI (\textit{r} = -0.785, \textit{p} = 0.037). Only strong negative correlations were found in the theta band. Specifically, there was a significant difference between the picture memory and frontal-occipital wPLI (\textit{r} = -0.821, \textit{p} = 0.024) and there was a significant difference between location memory and brain regions (temporal wPLI: \textit{r} = -0.968, \textit{p} $<$ 0.001; parietal wPLI: \textit{r} = -0.851, \textit{p} = 0.015; parietal-occipital wPLI: \textit{r} = -0.831, \textit{p} = 0.021). In alpha band, a negative correlation between the word-pairs memory and the occipital wPLI (\textit{r} = -0.763, \textit{p} = 0.046) and a positive correlation between positive memory and parietal wPLI (\textit{r} = 0.854, \textit{p} = 0.001) were observed. The most significant association was found in the spindle band. As a result, the word-pairs memory and brain regions (central-parietal wPLI: \textit{r} = -0.790, \textit{p} = 0.035; parietal regions: \textit{r} = -0.774, \textit{p} = 0.041) showed a negative correlation. In addition, the positive correlation between picture memory and brain regions (frontal-pareital wPLI: \textit{r} = 0.761, \textit{p} = 0.047; frontal-occipital wPLI: \textit{r} = 0.778, \textit{p} = 0.040) and negative correlation with occipital wPLI (\textit{r} = -0.928, \textit{p} $<$ 0.001) were found. On the other hand, wPLI in the beta and gamma bands showed no significant correlation with memory recall.

\section{Discussion} %다시 한번 전체적으로 수정하기!

% In this study,
In this study, we investigated the relationship between unconsciousness and memory consolidation. In the memory performance, spindle PSD during unconsciousness mainly had a positive correlation, and delta, alpha, and spindle wPLI during unconsciousness had a negative correlation. In the memory recall before and after the nap, gamma PSD during unconsciousness showed a strong positive correlation, especially in the parietal and occipital regions. The differences in wPLI during memory recall showed both positive and negative correlations.

% A. Memory task performance 관련 - 왜 증가한 게 있고, 감소한 게 있는지??
Comparing memory performance before and after the nap, word-pairs memory significantly increased, picture memory did not change, and location memory rather decreased. The reason for improving word-pairs memory performance after the nap is related to memory consolidation. The semantically related word-pairs is potentially sensitive to changes in the hippocampus-dependent consolidation process \cite{tononi2014sleep}. Therefore, memory consolidation during unconsciousness has been positively affected and is thought to have increased word-pairs memory performance. On the other hand, picture memory has been demonstrated to be less dependent on memory consolidation as in the previous studies \cite{sterpenich2007sleep}. Finally, the reduction in the performance of location memory may be due to the combination of simple picture memory task and more complex location memory task, resulting in fewer valid items \cite{ladenbauer2016brain}.

% B. spindle의 역할: 왜 Spindle PSD에서는 양의 상관관계, Spindle wPLI는 음의 상관관계? (Performance) *추가*
We observed that spindle activity during unconsciousness is closely related to memory consolidation. It has already been reported that the spindle band is associated with memory \cite{kalafatovich2020neural}. In addition, sleep spindle is even being used as a physiological marker for maintaining unconsciousness \cite{loomis1935potential}. In the alpha band, many negative correlations with memory performance were found in wPLI during unconsciousness. It is thought that the alpha band is involved in the inhibition of brain activity that is not involved in mental activity \cite{gevins2012long}. In summary, memory consolidation represents the positive or negative relationship with unconsciousness according to the role of frequency.

% C. spatial feature - 낮잠 전 후의 공간적 차이(통계) (dream recall) *추가*
With regard to spatial information, the parietal-occipital regions during consciousness were mainly related to memory consolidation in both PSD and wPLI. These parietal-occipital regions are neural correlates of consciousness as a posterior hot zone \cite{sarasso2015consciousness, lee2019connectivity}. It is considered that a major feature of these regions is associated with memory consolidation. Therefore, these regional changes can be used to assess unconsciousness based on memory consolidation.

% We compared EEG features to investigate the neurophysiological differences caused by unconsciousness during memory recall. In specific, the unconsciousness over temporal, parietal, and occipital regions in PSD showed a prominent feature with memory consolidation.  

% D. Gamma (증가)의 역할: memory 및 unconsciousness에서  % 양의 상관관계(모든밴드) *추가*
%In the gamma band, a strong positive relationship showed in PSD during unconsciousness and memory recall. The power of gamma oscillations, which are closely connected with the role of unconsciousness in brain function and cognition.

% limitation
Some limitations should be noted when interpreting these findings. First, the number of participants was relatively small and there was no comparison group. Second, we analyzed only successful recall sessions. In future work, a comparative analysis of successful and forgotten trials can be needed. 

%conclusion
In conclusion, we investigated the relationship between unconsciousness and memory consolidation. To assess unconsciousness, we performed two memory tasks before and after the nap. As a result, significant correlations with unconsciousness were explored in memory performance and differences between immediate and delayed memory recall sessions. These findings could help provide new insights into the assessment of unconsciousness by exploring memory consolidation based on BMI.

%\section*{Acknowledgment}

\bibliographystyle{IEEEtran}
\bibliography{REFERENCE}

\end{document}